\documentclass[%
 aip,
 prb,%
 amsmath,amssymb,
 reprint,%
]{revtex4-1}
\usepackage{graphicx}
\usepackage{dcolumn}
\usepackage{bm}
\usepackage[mathlines]{lineno}
\begin{document}

\preprint{AIP/123-QED}

\title{Thickness dependence of magnetic properties of (Ga,Mn)As}

\author{O. Proselkov}
 \email{proselkov@ifpan.edu.pl}
 \affiliation{Institute of Physics, Polish Academy of Sciences, Warszawa, Poland}
\author{D. Sztenkiel}
 \affiliation{Institute of Physics, Polish Academy of Sciences, Warszawa, Poland}
\author{W. Stefanowicz}
 \affiliation{Institute of Physics, Polish Academy of Sciences, Warszawa, Poland}
\author{M. Aleszkiewicz }
 \affiliation{Institute of Physics, Polish Academy of Sciences, Warszawa, Poland}
\author{J. Sadowski}
 \affiliation{Institute of Physics, Polish Academy of Sciences, Warszawa, Poland}
\affiliation{MAX-lab, Lund University, Lund, Sweden }
 \author{T. Dietl}
 \affiliation{Institute of Physics, Polish Academy of Sciences, Warszawa, Poland}
 \affiliation{ Institute of Theoretical Physics, Faculty of Physics, University of Warsaw, Warszawa, Poland}
\author{M. Sawicki}
 \email{mikes@ifpan.edu.pl}
 \affiliation{Institute of Physics, Polish Academy of Sciences, Warszawa, Poland}

\date{\today}

\begin{abstract}
We report on a monotonic reduction of Curie temperature in dilute ferromagnetic semiconductor (Ga,Mn)As upon a well controlled chemical-etching/oxidizing thinning from 15~nm down to complete removal of the ferromagnetic response. The effect already starts at the very beginning of the thinning process and is accompanied by the spin reorientation transition of the in-plane uniaxial anisotropy. We postulate that a negative gradient along the growth direction of self-compensating defects (Mn interstitial) and the presence of surface donor traps gives quantitative account on these effects within the p--d mean field Zener model with adequate modifications to take a nonuniform  distribution of holes and Mn cations into account. The described here effects are of practical importance for employing thin and ultrathin layers of (Ga,Mn)As or relative compounds in concept spintronics devices, like resonant tunneling devices in particular.
\end{abstract}


\maketitle


Dilute ferromagnetic semiconductors (DFS), such as (Ga,Mn)As, are extensively studied in search for new spintronic phenomena and towards potential applications in memory and information processing technologies.\cite{Ohno:2010_NM}
Among them, the most application-promising is the isothermal control of the magnetic phase\cite{Ohno:2000_N,Sawicki:2010_NP} and magnetic anisotropy change\cite{Chiba:2008_N} by the externally applied electric field, which recently led to application-viable demonstration of the electric field induced magnetization switching in sub-nanometer thin Fe-Co layers.\cite{Shiota:2012_NM}
Thus, there is a general interest in studying various (new) device configurations in which ultrathin, less than few nanometers, semiconducting (III,Mn)V layers are incorporated.
Therefore it is timely to provide experimental information how micromagnetism in such thin DFS compares with that of thicker layers for which a great deal of information has already been acquired.\cite{Sawicki:2006_JMMM,Jungwirth:2005_PRBb,Dietl:2010_NM}

In this letter we investigate changes of Curie temperature ($T_{\mathrm{C}}$) and magnetic anisotropy associated with systematic thinning of $d=15$~nm thick as-grown (Ga,Mn)As layers till the complete loss of the ferromagnetic signatures.
The magnitude of $T_{\mathrm{C}}$  is as high as $\sim 110$~K, which points to high sample quality, in particular, to a low density of antisite compensating donors.\cite{Myers:2006_PRB}
We establish the presence of two thickness regimes, for which the evolution of magnetic properties is determined by two distinct mechanisms. The first  is associated with the presence of a gradient in the concentration of interstitial Mn ions.
The second, operating at $d \lesssim 4$~nm,  is brought about by surface defects pining the Fermi energy in the mid gap region of GaAs, and thus depleting holes.\cite{Sawicki:2010_NP,Fujii:2011_PRL}
Our results  demonstrate, therefore, why low temperature annealing not only reduces the concentration of interstitial Mn but also "homogenizes" magnetic properties of (Ga,Mn)As, as observed in neutron studies.\cite{Kirby:2004_PRB}
Furthermore, the experimentally determined dependence $T_{\mathrm{C}}(d)$ in the second regime substantiates the previous theoretical model of ferromagnetism in interfacial space charge layers of (Ga,Mn)As.\cite{Sawicki:2010_NP,Nishitani:2010_PRB}
We also evidence a thickness induced 90$^{\circ}$ rotation ([110] $\Leftrightarrow$ [$\bar{1}$10]) of an in-plane uniaxial easy axis [a spin reorientation transition (SRT)].

Two $d = 15$~nm (Ga,Mn)As films (A and B) have been deposited at ~200$^{\circ}$C by low-temperature (LT) molecular beam epitaxy on (GaAs) on (100) substrates buffered by 500~nm thick LT--GaAs with a use of arsenic valved cracker effusion cell.
The concentration of substitutional Mn at Ga sites $x_{\text{sub}} \simeq 6$\% is estimated by the growth rate increase of (Ga,Mn)As in comparison to the GaAs buffer.\cite{Sadowski:2000_JVST}
The samples have not undergone any post-growth heat treatment.
The thickness dependent data are obtained by the controlled thinning of the magnetic films via sequential open air oxidation of the superficial part of the layer which had its native oxide removed beforehand by 30~s dipping in concentrated ($\approx 30$\%) HCl.\cite{Edmonds:2005_PRB,Olejnik:2008_PRB}
As this native oxide restores on an expense of the top-most part of the semiconductor film, multiple repetition of this method allows for very fine and uniform thinning of even macroscopically large areas.
The thickness of the reformed oxide depends on the oxidation time,\cite{Olejnik:2008_PRB} and it takes 26 etching-oxidation steps to completely remove ferromagnetic signal from the layer A which oxidized on open air for 6-9 hours between etching and magnetic measurements but only 13 steps are needed for the control sample B which oxidized about 24 hours.
As the observed reduction of $T_{\mathrm{C}}$ to thinning is essentially the same for both samples, we narrow this report to the main sample A only.

\begin{figure}
\includegraphics[width=8.3cm]{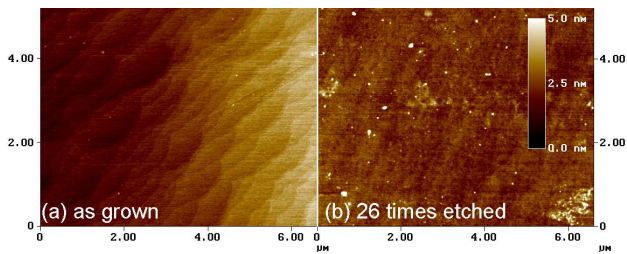}
\caption{\label{fig:AFM} (Color online) AFM images for two pieces from the same wafer (a) AFM image for as grown sample (b) AFM image for 26 times etched sample.}
\end{figure}

Atomic force microscopy (AFM) images  reassure us that the whole process proceeds in a truly planar fashion.
As indicated in Fig.~\ref{fig:AFM} even after 26 etchings  we do not find any macroscopic or submicrometer sized features suggestive a presence of a statistically relevant number of (Ga,Mn)As islands left on the GaAs substrate.
Neither do we observe any unidirectional surface undulation recently reported for similar (Ga,Mn)As layers.\cite{Piano:2011_APL}
Finally, we note that it is indeed oxygen needed to conduct the process, as no changes of signal are recorded in time domain during lengthy magnetic measurements which are done in a chemically inert helium atmosphere.

Magnetometry measurements are carried out on a home made superconducting quantum interference device (SQUID) magnetometer operating down to 5~K and up to 5~kOe.
Routinely field cooled (FC) measurements at the magnetic field $H =1$~kOe  are performed to assess the temperature dependence of the spontaneous moment $m(T)$ and to establish the magnitude of the saturation moment $m_{\mathrm{S}} \equiv m(1$~kOe, 5~K).
We find such an assessment of $m_{\text{S}}$ quite satisfactory, as prior to the whole experiment we checked at 5~K that the moment of the layers saturates already below 1~kOe for both major in-plane crystallographic directions, [110] and [$\bar{1}$10].
To establish $T_{\mathrm{C}}$ of the layer every FC measurement is followed by a thermoremnant (TRM) one carried out on increasing temperature at $H=0$ until the remnant moment vanishes completely.
This set of measurements is performed twice for these two in-plane directions,  and is repeated after \textsl{every} step of thinning.
We follow the experimental code described recently in Ref.~\onlinecite{Sawicki:2011_SST}.

\begin{figure}[t]
\includegraphics[width=8.5cm]{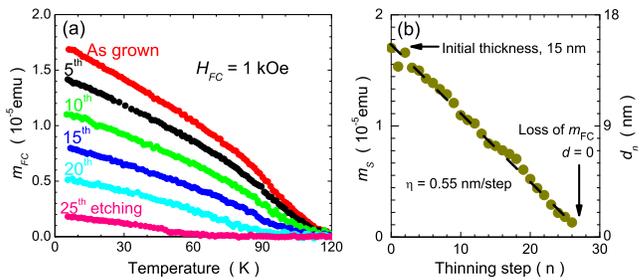}
\caption{\label{fig:mSvsEtching} (Color online) (a) Temperature dependence of the field cooled moment $m_{\mathrm{FC}}$ of layer A measured after every 5th stage of the thinning process. (b) Reduction of the saturation moment [taken as $m_{\mathrm{FC}}(5$~K)] versus number of the thinning step. Dashed line defines the thinning efficiency $\eta$. Calculated layer thickness of the remaining material, $d_{\text{n}} = 15-n\eta$, is given on the right $y$-axis.}
\end{figure}

Figure~\ref{fig:mSvsEtching}a  exemplifies $m(T)$ collected at every 5th step of thinning which evidence a gradual weakening of the ferromagnetic response in terms of reduction of both the magnitude of magnetic moment (and so of its saturation value) and its onset temperature ($\sim T_{\mathrm{C}}$).
As in the carrier mediated ferromagnetism\cite{Dietl:2001_PRB} $m_{\text{S}}$ depends primarily on the \emph{number} of these Mn cations which are bound ferromagnetically by holes it is straightforward to assume that the drop of $m_{\text{S}}$ is caused essentially by the reduction of the volume of the layer.
Therefore, and on the account of the AFM studies, we solely assign this drop of $m$ to the step-by-step reduction of the (Ga,Mn)As thickness and employ $m_{\text{S}}$ to size the thickness of the layer after each thinning stage.

In Fig.~\ref{fig:mSvsEtching}b we plot the values of $m_{\text{S}}$ versus the number of the thinning step $n$.
The observed there remarkably linear dependence allows us to establish the average thickness loss per each thinning step ($\eta \simeq 0.55$~nm), which is the thickness of the native oxide forming on the free surface of the (Ga,Mn)As between etching and beginning of measurement,
and calculate (Ga,Mn)As thickness $d_{\text{n}}=15-n\eta$~nm left after each stage of the thinning process.
We note in parenthesis that the same experimental procedure gives larger $\eta \simeq 1$~nm for sample B, and, given the different time of oxidation, both values agree with those reported in Refs.~\onlinecite{Olejnik:2008_PRB} and \onlinecite{Horak:2011_PRB}.

\begin{figure}[t]
\includegraphics[width=8.5cm]{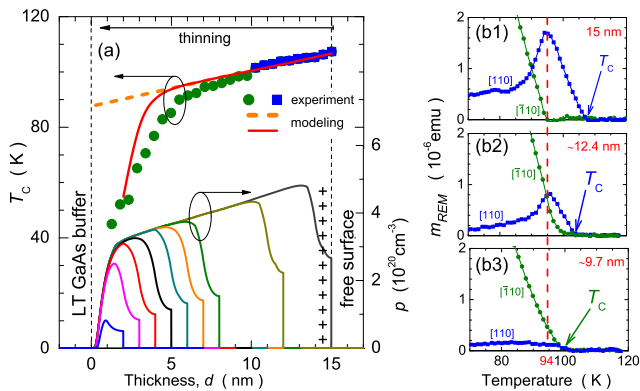}
\caption{\label{fig:MainRes} (Color online)  (a/top) (Full points) experimentally determined values and (dashed and solid lines) results of modeling of Curie temperature $T_{\mathrm{C}}$ in 15~nm (Ga,Mn)As as a function of decreasing layer thickness $d$.  (a/bottom)  Hole profiles for various values of $d$ calculated by solving the Poisson equation with the negative gradient of the concentration of Mn at interstitial sites and electrostatic effects present at the top and bottom boundaries of the layer (see the text). (b1-b3) Squares -- [110], bullets -- $[\overline{1}10]$ projection of the thermoremnant moment at elevated temperatures recorded for: (b1) as grown layer, (b2) after 5th step, and (b3) after 10th step of thinning. The corresponding layer thickness is quoted. The vertical dashed line marks an arbitrary selected temperature $T = 94$~K at which thickness-driven spin reorientation transition is observed. Arrows indicate $T_{\mathrm{C}}$ for these cases.}
\end{figure}

Bullets in Fig.~\ref{fig:MainRes}a illustrate the main effect observed in this study:
a continuous decrease of $T_{\mathrm{C}}$ in response to the reduction of the layer thickness, and, quite remarkably, that the whole process is already effective at the very first steps of thinning, that is for $d \simeq 15$~nm.
We further note, that the whole observed $107 \rightarrow 43$~K drop of $T_{\mathrm{C}}$ consists of two parts suggesting that two different mechanisms conspire.
According to the previous findings,\cite{Sawicki:2010_NP,Nishitani:2010_PRB} we expect that the fast drop of $T_{\mathrm{C}}$ for the lowest thicknesses results from a significant reduction of hole concentration $p$ due to the depletion zones present at both boundaries of the layer.
However, in order to account for the initial reduction of $T_{\mathrm{C}}$ ($d =15 \rightarrow 5$~nm) we need to assume an existence of a positive gradient (counting along $z$) of either $x_{\text{eff}}$ or $p$.

Within the framework of the p--d Zener model, $T_{\mathrm{C}}$ depends independently on $x_{\text{eff}}$ and $p$, so an adequately strong positive gradient of any of these two quantities would explain this behavior.
However, we actually need a simultaneous reduction of both $x_{\text{eff}}$ and $p$ to take place on thinning in order to explain simultaneously occurring SRT, which is  heralded by $[110] \rightarrow [\overline{1}10]$ rotation of the magnetic easy axis taking place at \emph{constant} temperature,\cite{Sawicki:2005_PRB,Zemen:2009_PRB,Stefanowicz:2010_PRBb} as exemplified for $T = 94$~K in panels b1--b3 of Fig.~\ref{fig:MainRes}.
We underline here that the presence of SRT calls for the determination of $T_{\mathrm{C}}$ initially from [110] projection of TRM (for $d > 10$~nm) and from the $[\overline{1}10]$ projection afterwards, as it is differentiated in Fig.~\ref{fig:MainRes}a by use of two different colors and symbols to mark the experimental values of $T_{\mathrm{C}}$.

We now show that the presence of a negative gradient of $x_{\text{I}}$ in the layer is a sufficient requirement to obtain the requested changes of $x_{\text{eff}}$ and $p$ on $z$ and we use the p--d Zener model of ferromagnetism in DFS to give a numerical account on the scale of the effect.
Firstly, we fix $x_{\text{sub}}$ as depth independent according to the stability of the reflection high-energy electron diffraction pattern recorded during the growth.
We assume then, for the sake of simplicity, a linear decrease $x_{\text{I}}$ on $z$, $x_{\text{I}}(z)=x_{\text{I}}(15) + (15-z)b$, $b > 0$, and calculate accordingly $x_{\text{eff}}(z)$ and $p(z)$ using  Fig.~8 of Ref.~\onlinecite{Jungwirth:2005_PRBb} to evaluate $x_{\text{I}}(15)$.
Now with the problem reduced to just only two free parameters, namely:  $x_{\text{sub}}$ and the gradient magnitude $b$, their values are obtained by fitting calculated $T_{\mathrm{C}}(d)$ to the experimental data using the following formulae developed to describe $T_{\mathrm{C}}$ of thin and nonuniform layers of (Ga,Mn)As:\cite{Sawicki:2010_NP,Nishitani:2010_PRB}
\begin{equation}
T_{\mathrm{C}}=\int dz T_{\mathrm{C}}^{3D}\left[ p(z),x_{\text{eff}}(z)\right] \int dz \frac{p^{2}(z)}{p_s^2},
\label{eq:wzorTD}
\end{equation}
where, due to a rather short phase coherence of holes at these temperatures $L_{\phi} \approx 1$~nm,\cite{Sawicki:2010_NP,Nishitani:2010_PRB}  $z$ runs  from 0 to $d$ in the intervals of 1~nm and the maximal obtained value is assign to current  $T_{\mathrm{C}}(d)$.\footnote{We confirm here that the obtained results depend marginally on the choice of $L_{\phi}$ up to 4 nm.}
Here $T_{\mathrm{C}}^{3D}$ is the Curie temperature calculated within the conventional p--d Zener model\cite{Dietl:2001_PRB} at given $z$, $p_s = \int dz p(z)$ is the sheet hole density.
Indeed, as indicated in Fig.~\ref{fig:MainRes}a/top by the dashed line, the presented above simple model reproduces the data remarkably well.
The fit yields $x_{\text{sub}}=5.7$\%, the value consistent with that established from the layer growth rate, and $b = 0.023$~\%/nm, or $x_{\text{I}}$ drops from 2.15 at the beginning of the growth to 1.8\% at the end.
Importantly, the total Mn concentration changes along these 15~nm rather marginally from 7.8 to 7.45\% (a relative drop by merely $\sim$5\%), what is currently beyond the resolution of even the most advanced \emph{direct} atomic concentration profilers like secondary ion mass spectroscopy and  three-dimensional atomic probe.\cite{Prv_Comm}
Conversely, we can state that, despite being tedious, this method provides the most accurate (indirect) assessment of the depth dependence of the two most numerous  Mn species in very thin (Ga,Mn)As layers.

At the final stage we add electrostatic effects due to the presence of antisite As$_{\text{Ga}}$ donors of concentration $N_{\text{D}}$ in the
LT--GaAs buffer adjacent to (Ga,Mn)As channel and donor-like traps at the free (Ga,Mn)As surface.
We model the traps by introducing an ever-present topmost 1~nm region of the layer containing $N_{\text{I}}$ donors.
We pin the Fermi energy at the midgap of GaAs substrate residing 50 nm below the bottom of our layer.
Then we solve the Poisson equation within next\textbf{nano}$^3$ package\cite{Birner:2006_APP} for the established already distribution of interstitial double donors $x_{\text{I}}(z)$ and look for a set of $N_{\text{I}}$ and $N_{\text{D}}$ that reproduces the experimental $T_{\mathrm{C}}(d)$.

Figure \ref{fig:MainRes}a shows the hole distribution profiles for various thicknesses (bottom part) and (top part) the corresponding $T_{\mathrm{C}}$ values calculated at particular channel thickness (thick line) for $N_{\text{I}}=2.8 \times 10^{20}$~cm$^{-3}$ and $N_{\text{D}}= 5 \times10^{19}$~cm$^{-3}$.
Although this is not a rigorous fit and the results depend to some extent on the magnitudes of these two adjustable concentrations, we can conclude that for generally similar values of $N_{\text{I}}$ and $N_{\text{D}}$ to those reported previously by some of us\cite{Sawicki:2010_NP,Nishitani:2010_PRB} the presented here model describe \emph{quantitatively} the magnitude of $T_{\mathrm{C}}$ changes on thinning.

We are now in a position to address the question why there exists the Mn$_{\text{I}}$ gradient in (Ga,Mn)As, and perhaps other (III,Mn)V, at the first place.
We argue that this is this dense and narrow pocket of the surface donor-like states that influences how Mn$_{\text{I}}$, the by far dominating self-compensating defect in (Ga,Mn)As, get distributed during the growth.
We recall here that both the traps and Mn$_{\text{I}}$ are having a similar densities and are positively charged, so they repel each other.
But as the traps are fixed to the surface and Mn$_{\text{I}}$ are quite mobile at the growth temperatures (which is why the LT annealing does work in these compounds), so the Mn$_{\text{I}}$ are being constantly pushed back from growth front towards the substrate.
It is beyond the scope of this letter to provide with a selfconsistent solution of the full thermodynamics of the process, but we do see that the combined push backwards by the surface positive charge and simultaneous push forward from the already accumulated Mn$_{\text{I}}$ at the deeper parts of the already grown layer may/should result in an equilibrium \emph{negative} gradient of Mn$_{\text{I}}$, which gets immediately frozen down once the growth is completed, temperature reduced and Mn$_{\text{I}}$ immobilized.
Therefore, we believe that apart from many possible technical-related reasons the proposed here mechanism should be effective in all (Ga,Mn)As layers with the gradient coefficient $b$ being a decreasing function of the thickness, except, perhaps of hydrogen-codoped layers where virtually no Mn$_{\text{I}}$ are expected to form.
Similarly, in materials like (Ga,Mn)Sb where the surface states are filled by the band states there should be no 'push-back' effect and so both $x_{\text{I}}$ and $p$ should stay constant along the depth of the layer, unless a variation of growth parameters tells the system otherwise.

\begin{figure}[t]
\includegraphics[width=8cm]{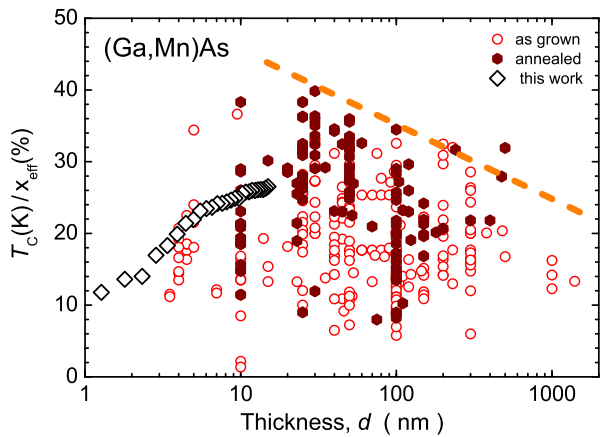}
\caption{\label{fig:Tc-d} (Color online) Literature data on $T_{\mathrm{C}}$ (given in units of the effective Mn contraction, $x_{\text{eff}}$) of (Ga,Mn)As plotted as a function of thickness $d$. Open symbols - as grown, bullets - after annealing, diamonds - present data. Where not explicitly stated, the $x_{\text{eff}}$ values have been calculated according to Ref.~\onlinecite{Jungwirth:2005_PRBb}. Essentially the same picture is obtained if $T_{\mathrm{C}}$ alone is plotted.\cite{Suppl_Info} 
The thick dashed line is a guide for the eye and indicates the main trend of improvement of $T_{\mathrm{C}}$ on lowering $d$.}
\end{figure}

It is interesting to compare our results to previously determined values of $T_{\mathrm{C}}$ in films of various thicknesses $d$.
According to data collected in Fig.~\ref{fig:Tc-d}, {\em annealed} samples tend to show  increasing $T_{\mathrm{C}}$ values when $d$ decreases down to 20~nm. This  trend reflects a decreasing efficiency of the low temperature annealing with the layer thickness, the effect expected within the model of interstitial Mn diffusion.\cite{Edmonds:2004_PRL}
As already discussed, our results reveals an opposite trend in as-grown samples.

In conclusion, we have performed systematic studies how $T_{\mathrm{C}}$ in (Ga,Mn)As depends on thickness, finding that the observed nearly 60\% drop in the magnitude of $T_{\mathrm{C}}$ begins already in 15~nm layers where no depletion due to surface donor defects is expected.
We have assigned this effect to a build--in negative gradient of the Mn interstitials' concentration and successfully reproduced the experimental finding in the frame of the adequately modified p--d Zener model of ferromagnetism to the case of nonuniform hole and Mn distributions.
This result is expected to help to understand the properties and behavior of various spintronic devices which rely they functionalities on thin or even ultra thin layers of (Ga,Mn)As and possibly other DFS.

The work was supported in part by the European Research Council through the FunDMS Advanced Grant within the "Ideas" 7th Framework Programme of the EC and EC Network SemiSpinNet (PITN-GA-2008-215368).



%

\vspace{3cm}

\Large{\textbf{Supplementary information}}

\normalsize

\begin{figure}[bh]
\includegraphics[width=8.5cm]{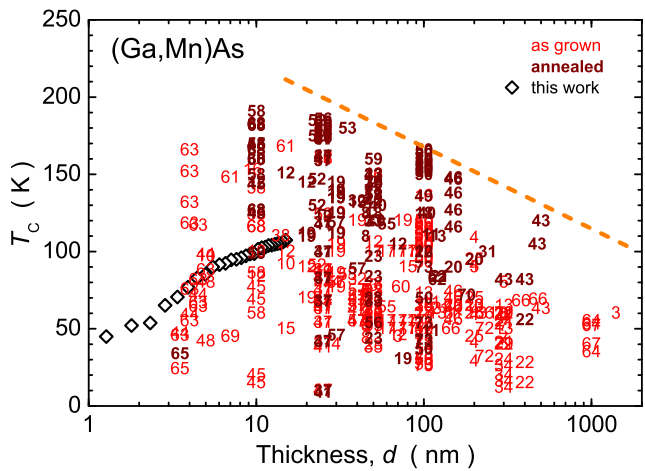}
\caption{\label{fig:SIFig}  Literature data on $T_{\mathrm{C}}$ of (Ga,Mn)As plotted as a function of thickness $d$. Red numbers - as grown, bold brown numbers - after annealing, diamonds - present data.
The thick dashed line is a guide for the eye and indicates the main trend of improvement of $T_{\mathrm{C}}$ on lowering $d$.}
\end{figure}

Figure \ref{fig:SIFig} shows the same $T_{\mathrm{C}}$ values as those in the Fig.~4 of the main paper, but plotted without normalization with respect to the effective Mn concentration  $x_{\text{eff}} = x_{\text{sub}} - x_{\text{I}}(z)$, where $x_{\text{sub}}$ and $x_{\text{I}}$ are concentrations of Mn ions on Ga and interstitial sites, respectively.

Both figures were constructed form data taken from the publications listed below (ordered according to the year of appearance).


%

\end{document}